\pdfoutput=1

\documentclass{article}

\usepackage[final]{neurips_2021}

\makeatletter\let\@noticestring\relax

\usepackage[utf8]{inputenc}
\usepackage[T1]{fontenc}
\usepackage{hyperref}
\usepackage{url}
\usepackage{booktabs}
\usepackage{amsfonts}
\usepackage{nicefrac}
\usepackage{microtype}
\usepackage{xcolor}
\makeatletter\let\NAT@force@numbers\relax\makeatother
\title{IMDB-WIKI-SbS: An Evaluation Dataset for Crowdsourced Pairwise Comparisons}

\author{%
  Nikita Pavlichenko \\
  Yandex\\
  Moscow, Russia \\
  \texttt{pavlichenko@yandex-team.ru} \\
  \And
  Dmitry Ustalov \\
  Yandex \\
  Saint Petersburg, Russia \\
  \texttt{dustalov@yandex-team.ru} \\
}

\begin{document}

\maketitle

\begin{abstract}
Today, comprehensive evaluation of large-scale machine learning models is possible thanks to the open datasets produced using crowdsourcing, such as SQuAD, MS~COCO, ImageNet, SuperGLUE, etc. These datasets capture objective responses, assuming the single correct answer, which does not allow to capture the subjective human perception. In turn, pairwise comparison tasks, in which one has to choose between only two options, allow taking peoples' preferences into account for very challenging artificial intelligence tasks, such as information retrieval and recommender system evaluation. Unfortunately, the available datasets are either small or proprietary, slowing down progress in gathering better feedback from human users. In this paper, we present IMDB-WIKI-SbS, a new large-scale dataset for evaluating pairwise comparisons.\footnote{\url{https://github.com/Toloka/IMDB-WIKI-SbS}} It contains 9,150 images appearing in 250,249 pairs annotated on a crowdsourcing platform. Our dataset has balanced distributions of age and gender using the well-known IMDB-WIKI dataset as ground truth. We describe how our dataset is built and then compare several baseline methods, indicating its suitability for model evaluation.
\end{abstract}

\section{Introduction}

Crowdsourcing has become an essential choice for producing large-scale training and evaluation datasets for a broad spectrum of machine learning tasks, including computer vision, natural language processing, etc. However, crowdsourcing requires a careful task design that is usually reduced to a classification task with a single known objective response~\citep{Zheng:17}. In such applications as information retrieval and recommender systems training and evaluation, there is no objective response as the obtained rankings are relatively complex to evaluate properly. This challenge is usually addressed by relaxing the requirement of the objective answer and trying to gather the \emph{subjective human opinions} on how the objects should be ranked~\citep{Beutel:19}.

As most humans are non-experts in ranking evaluation, preferences are gathered using pairwise comparisons (side-by-side, or SbS comparisons). In this scenario, the crowdsourcing tasks offered to the workers show a pair of objects with a question of which object is better. However, to train or evaluate the systems described above, it is necessary to obtain the ranking from the resulting comparisons. This process is called the \emph{aggregation} and is fraught with various difficulties such as dealing with noisy or unreliable responses from crowd workers. Scientific community has developed specific models for aggregation of pairwise comparisons into ranked lists, including Bradley-Terry~\citep{Bradley:52}, CrowdBT~\citep{Chen:13}, factorBT~\citep{Bugakova:19}, and others~\citep{Zhang:16}. The resulting lists are then used as ground truth datasets for computing quality criteria, such as normalized discounted cumulative gain or expected reciprocal rate. Unfortunately, most evaluation datasets reported in studies are either small but open~\citep{Janssens:12,Chen:13,Zhang:16} or large but proprietary~\citep{Bugakova:19}, which complicates systematic studies of these algorithms and benchmarking models, forcing the researchers to use non-realistic synthetic models in their experiments.

\begin{table}[htbp]
\centering
\caption{\label{tab:datasets}Summaries for related datasets and the dataset presented in this paper.}
\begin{tabular}{lrr}\toprule
\textbf{Dataset} & \textbf{\# of Objects} & \textbf{\# of Pairs} \\\midrule
PeopleAge~\citep{Janssens:12,Zhang:16} & 50 & 1,225 \\
PeopleNum~\citep{Khan:14} & 39 & 741 \\
EventTime~\citep{Zhang:16} & 100 & 4,950 \\
ImageClarity~\citep{Zhang:16} & 100 & 4,950 \\
Graded Readability~\citep{Chen:13} & 491 & 13,856 \\\midrule
IMDB-WIKI-SbS [This Paper] & 9,150 & 250,249 \\\bottomrule
\end{tabular}
\end{table}

We attempt in this paper to bridge the gap between these two extremes, presenting IMDB-WIKI-SbS, a new large-scale dataset for evaluation pairwise comparisons, building on the success of a well-known benchmark for computer vision systems IMDB-WIKI~\citep{Rothe:18}.\footnote{\url{https://data.vision.ee.ethz.ch/cvl/rrothe/imdb-wiki/}} Our dataset uses the age information offered by IMDB-WIKI as ground truth while providing a balanced distribution of ages and genders of people in photos. Table~\ref{tab:datasets} shows the comparison of our dataset to other datasets used in research papers in terms of size.

The rest of the paper is organized as follows. Section~\ref{sec:dataset} explains how IMDB-WIKI-SbS was derived and sampled from the original IMDB-WIKI dataset. Section~\ref{sec:annotation} describes how the sampled pairs of photos were annotated on the Toloka crowdsourcing platform. Section~\ref{sec:evaluation} presents the empirical evaluation of three baselines on our new dataset. Section~\ref{sec:conclusion} concludes with final remarks.

\section{\label{sec:dataset}Dataset}

This section describes how we chose the photos for our dataset from a popular IMDB-WIKI dataset~\citep{Rothe:18} and produced sample pairs for further crowdsourced annotation.

\subsection{Choice of Photos}

We have derived our dataset from the IMDB-WIKI as follows. As in the original dataset, the odds ratio between males and females is roughly 1.4:1, and 50\% of the data represent people aged 28--45, which does not allow for unbiased comparison between age and gender groups provided in the dataset. We decided to narrow our dataset span from 10 to 70 years. We excluded all the entries in which the face bounding box and gender label are missing. Then, we sampled uniformly 75 photos per age and gender. One person can appear only once in every group, but the same person can appear in multiple groups. If a person has multiple photos in the group, we have chosen the one with the highest face detector score provided by the dataset authors. As the result, our final dataset contained $(60 + 1) \times 2 \times 75 = 9,\!150$ photos. We have noticed that the original IMDB-WIKI contains the wrong gender information for some photos. However, we believe the amount of such mistakes does not affect the evaluation, given the size of our dataset.

\subsection{Sampling Pairs}

To produce a dataset allowing algorithm benchmarking, we decided to provide a \emph{connected} graph of pairwise comparisons. To do so, we generated an Erd\H{o}s-R\'{e}nyi graph containing 250,249 edges using the NetworkX toolkit~\citep{Hagberg:08}. When generating the graph, we explicitly shuffled the pairs to ensure that the distribution of labels would be uniform during crowd annotation. Although pairwise comparisons graphs might be disconnected in real-world scenarios, we believe that it is simpler to remove the available edges from our dataset and then evaluate how well the algorithm recovers and handles the missing ones. We decided to focus on the connected version due to the evaluation of the popular Bradley-Terry model that requires such a graph configuration~\citep{Bradley:52}.

\section{\label{sec:annotation}Annotation}

This section describes how we set up the quality control and annotated our dataset on the Toloka crowdsourcing platform, \url{https://toloka.ai/}.

\subsection{Quality Control}

We used the following quality control configuration. We limited the minimum time for task completion to at least four seconds. If the worker skipped more than ten task pages, we blocked them from accessing additional tasks. Our tasks were available to the top-50\% workers on Toloka who set English as the spoken language. Additionally, to avoid malicious behavior, we used 200 golden comparisons that contained images of people with more than 20 years gap in age. We mixed these golden comparisons with the regular tasks. Failing to solve them correctly led to blocking access to our tasks.

\subsection{Data Annotation}

This dataset contains images of people with reliable ground-truth age assigned to every image. The task interface contained two images and two buttons, left and right. We intentionally omitted the ``no answer'' to avoid ambiguous cases not handled by all the algorithms and prevent abusing this option by malicious workers. This setup allowed us to obtain 250,249 comparisons by 4,091 workers. One pair was annotated by only one unique worker, but a worker could annotate multiple tasks. Each task page contained three regular tasks and one control task in random order. While annotating, we used the cropped face images instead of the complete photos. On average, each task page was done in 23 seconds, and there were about 20 task pages per worker. The whole annotation process took 6 hours. As a result, 125,302 pairs labeled ``left'' and 124,947 pairs labeled ``right''.

\section{\label{sec:evaluation}Evaluation}

This section describes how we perform basic evaluation of baseline methods on our dataset using the commonly used normalized discounted cumulative gain (NDCG@k) score~\citep{Jarvelin:02}.

\subsection{Baselines}

We have evaluated the performance of four baselines: a specific algorithm for aggregation of pairwise comparisons, an algorithm that utilizes only comparisons graph information, a random baseline, and the reversed ground-truth ranking. We provide a brief description of these methods below.

\paragraph{Bradley-Terry.} Bradley-Terry is a probabilistic model for predicting the outcome of pairwise comparisons~\citep{Bradley:52}. Under this model, each item $i$ is associated with some positive real-valued score $p_i$. Given two items $i$ and $j$ with scores $p_i$ and $p_j$ respectively, probability of $i$ to be ranked higher than $j$ is $\frac{p_i}{p_i + p_j}$. The likelihood of received comparisons according to this probabilistic model is maximized with respect to the scores. We use the implementation of the Bradley-Terry model available in the Crowd-Kit computational quality control toolkit for crowdsourcing~\citep{Ustalov:21}.

\paragraph{PageRank.} PageRank is a classical algorithm for computing node centralities in a directed graph~\citep{Brin:98}. We built our directed graph as follows. As nodes, we used all the 9,150 images in our dataset. As arcs, we used all the 250,249 pairs in our dataset. The direction of an arc represents which image in the pair was labeled as older during the annotation. The resulting ranking is obtained by sorting the items by their PageRank scores in descending order.

\paragraph{Random.} As a random baseline, we used a ranking obtained by sorting the items by a uniformly generated random score. We ran this baseline to show the performance gain obtained by specialized methods.

\paragraph{Reversed GT.} As a sanity check, we used a reversed ground truth (GT) dataset to ensure the methods offer reasonable ranking order.

\begin{table}[htbp]
\centering
\caption{\label{tab:pairwise}Comparison of pairwise aggregation methods on the whole IMDB-WIKI-SbS dataset.}
\begin{tabular}{lrrr}\toprule
\textbf{Method} & \textbf{NDCG@100} & \textbf{NDCG@100 (Female Only)} & \textbf{NDCG@100 (Male Only)} \\\midrule
Bradley-Terry & $0.859$ & $0.879$ & $0.864$ \\
PageRank & $0.803$ & $0.856$ & $0.864$ \\
Random & $0.490$ & $0.533$ & $0.475$ \\
Reversed GT & $\approx 0$ & $\approx 0$ & $\approx 0$ \\ \bottomrule
\end{tabular}
\end{table}

\subsection{Results}

Our evaluation in Table~\ref{tab:pairwise} shows that even the simple pairwise comparison aggregation model performs well on our dataset compared to the random baseline while leaving space for the more principled models. Additionally, we evaluated the same baselines on two equally sized parts of our datasets separated by gender and found no specific difference in algorithm performance. To assess the difficulty of each age group, we evaluated the algorithm performance among different age groups in Table~\ref{tab:ages}. Even though each group contained the same number of photos, aggregation algorithms performed worse than the random baseline. We believe there are two reasons for this. First, for workers, between-group comparisons were more straightforward to interpret than the in-group ones, leading to the different grades of difficulty to the algorithms and explaining the dramatic difference to the evaluation on the whole dataset. Second, as the comparison graph, in this case, is not connected, the Bradley-Terry model failed to deliver a meaningful ranking.

\begin{table}[htbp]
\centering
\caption{\label{tab:ages}Comparison on pairwise aggregation methods on different age groups in IMDB-WIKI-SbS. NDCG@10 is used due to the relatively small number of objects per group.}
\begin{tabular}{lrrrrrr}\toprule
\textbf{Method} & \textbf{[10; 20]} & \textbf{[20; 30]} & \textbf{[30; 40]} & \textbf{[40; 50]} & \textbf{[50; 60]} & \textbf{[60; 70]} \\\midrule
Bradley-Terry & $0.358$ & $0.461$ & $0.622$ & $0.436$ & $0.681$ & $0.570$ \\
PageRank & $0.352$ & $0.682$ & $0.528$ & $0.629$ & $0.564$ & $0.715$ \\
Random & $0.507$ & $0.501$ & $0.547$ & $0.517$ & $0.549$ & $0.577$ \\\bottomrule
\end{tabular}
\end{table}

\section{\label{sec:conclusion}Conclusion}

In this work, we have collected the largest available open dataset for crowdsourced pairwise comparisons. The described IMDB-WIKI-SbS dataset has been published at \url{https://github.com/Toloka/IMDB-WIKI-SbS}, and its labels are made available under a CC~BY license. We believe this dataset will foster the development of better human-computer systems by allowing careful evaluation of these systems against human judgments and encourage the scientific community to develop better aggregation methods for crowdsourced pairwise comparisons.

\bibliographystyle{plain}
\bibliography{imdb-wiki-sbs}

\end{document}